# Fluid-structure interaction analysis of airflow in the lung alveolus

M. Monjezi[1], A. Rouhollahi[2], M. S. Saidi[3], B. M. Afshary[4]

[1]Department of Mechanical Engineering, Sharif University of Technology, Tehran, Iran; m_monjezi@mech.sharif.edu
[2]Department of Biomedical Engineering, Iran University of Science and Technology; rouhollahi@mecheng.iust.ac.ir
[3]Department of Mechanical Engineering, Sharif University of Technology, Tehran, Iran; mssaidi@sharif.edu
[4]Department of Mechanical Engineering, Sharif University of Technology, Tehran, Iran; bmafshary@yahoo.com

**Abstract**
The flow simulation in alveolar region is imperative in understanding transport of particles and designing aerosol drug delivery systems. Air flow is dependent on alveolar geometry and ventilation conditions. In this work a three dimensional honey-comb like geometry is constructed. A fluid-structure analysis is performed to study normal breathing airflow. Mechanical properties of alveolar wall tissue play an important role in the deformation of alveoli. We have used three distinct material models, linear, non-linear isotropic elastic and viscoelastic, in order to simulate the mechanical behavior of alveolar wall tissue and air flow simulation in it. Our simulation shows that linear and non-linear elastic models cannot describe alveolar tissue behavior as well as viscoelastic.

## Introduction

In contrast to the progress achieved in upper airway reconstruction, the success in three-dimensional (3D) imaging of alveolar region has been limited by its size and accessibility [1]. Recent studies have attempted to reconstruct the 3D image-based alveolar structure and define its dynamics. Tsuda et al. [2] performed simulations on simplified expanding and contracting torus surrounding a central channel. Later, Henry et al. [3] expanded this axi-symmetric model to 9-cell geometry.

Unsteady 3D simulation efforts of fluid flow in the alveolated ducts and sacs in the past decade have been scarce. Darquenne and Paiva [4] adopted a simplified model of the alveolated duct using sections of an annular ring around a central channel. Harrington et al. [5] used a similar representation, but compared the effect of acinar branching using a bifurcation model. Haber et al. [6] performed analytical investigation using a spherical cap to represent a single alveolus. Some of the later works also used self-similar breathing motion although using an isolated 3D cavity to represent an alveolus ([7] , [8]).

The mode of expansion and contraction of alveoli during normal respiration is not known. Various models in literature have considered rigid wall in their simulations. It has been shown by various researchers that major differences arise both in the flow structure and particle deposition characteristics when the motion of alveolar wall and alveolar duct are not properly included in the model. Hence, the models of acinus have to be supplemented with appropriate boundary conditions to mimic breathing. In the last decade various efforts have started incorporating alveolar wall motion into their analysis. Most of them use a self-similar wall motion with a prescribed sinusoidal function which could not illustrate real motion of alveolar wall. The direct relationship between tissue motion and airflow in the alveoli naturally suggests an FSI simulation technique. Dailey and Ghadiali have developed fluid-structure analysis in a 2D model of alveolar sacs. They have considered both elastic and Kelvin-Voigt viscoelastic models for alveolar tissue. In this paper a fluid-structure interaction model is performed to study normal breathing airflow in 3D honey-comb like geometry of alveoli. For mechanical properties of alveolar tissue we have used more realistic models according to experimental data previously reported.

## Geometry

We use a honey-comb like geometry [9] with a single entrance for representing terminal alveoli. A schematic of this 3D geometry is shown in Figure 1.

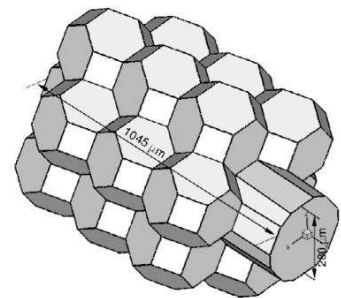

Figure 1: 3D alveolar sac geometry with a single entrance

For representing alveolar wall tissue, we have considered a 8 micron layer around the outer surface of geometry similar to Gefen et al.[10].

## Governing equations and boundary condition

Fluid domain
Fluid flow is governed by the incompressible continuity and Navier–Stokes equations.

$$\rho_f \frac{\partial v_i}{\partial x_i} = 0, \qquad (1)$$

$$\rho_f \frac{\partial v_i}{\partial t} + \rho_f v_j \frac{\partial v_i}{\partial x_j} = -\frac{\partial p}{\partial x_i} + \mu \frac{\partial^2 v_i}{\partial x_j \partial x_j}, \qquad (2)$$

where $\rho_f$ is the fluid density, $\mu$ is the fluid viscosity, $v_i$ is the velocity vector, $x_i$ is the position vector, t is time, and p is the fluid pressure.

Since in FSI fluid and structure deformations combine, the eulerian description of the fluid flow is no longer applicable. Therefore, the governing equations of fluid flow using an Arbitrary-Lagrangian-Eulerian (ALE) formulation are solved.

The boundary of the fluid domain is divided into the following regions for the assignment of boundary conditions: inlet, outlet, and the fluid-structure interaction interface. At the inlet, we don't apply any condition and let to be specified automatically with FSI solution. Also no-slip wall condition is applied at the outlet.

Structural domain

Tissue deformation in the solid domain is governed by momentum conservation equation given by Eq. (3). In contrast to the ALE formulation of the fluid equations, a lagrangian coordinate system is adopted.

$$\rho_s \frac{\partial^2 d_i^s}{\partial t^2} = \frac{\partial \sigma_{ij}^s}{\partial x_j}, \qquad (3)$$

Where $\rho_s$ is the solid density and $\sigma_{ij}^s$ stands for the solid Cauchy stress tensor.

The wall material implemented in this work represents a tissue of average characteristics consistent with actual mechanical behavior of alveolar wall tissue. As a first approximation, Linear and Non-Linear elastic materials are used similar to previous works [10, 11]. Young's modulus of 4 kPa and poisson's ratio of $\nu$=0.4 are used for Linear Elastic model.

The viscoelastic properties of the alveolar connective tissue are modeled using the quasi-linear theory [12]. The quasi-linear approach models the time-dependent stress– strain properties of the tissue as the product of a reduced relaxation function, a function of time alone, and an elastic stress response, a function of strain alone. The stress relaxation function defines the history of the stress response, the transient changes in stress with strain. The reduced relaxation function, G(t), is given by:

$$G(t) = 0.4 + 0.6(t+1)^{-0.5} \qquad (4)$$

G(t) refers to the time dependence of the stress response relative to a steady-state stress of the fiber at that strain. The stress response of a fiber bundle is given by Eq. 5.

$$\sigma(t) = \sigma_0[\varepsilon(t)] + \int_0^t \sigma_0[\varepsilon(t-\tau)] \frac{\partial G(\tau)}{\partial \tau} d\tau = \qquad (5)$$
$$\sigma_0[\varepsilon(t)] + H[\varepsilon(t),t]$$

where $\sigma(t)$ is the dynamic stress response, $\sigma_0[\varepsilon(t)]$ is the instantaneous stress response to a strain $\varepsilon(t)$, t is the time, and $H[\varepsilon(t),t]$ is a history integral.
we have considered $\sigma_0$ is not dependent on time and can be describe by Eq. 6 [10]. In this equation $\lambda$ is the stretch ratio.

$$\sigma_0 = \lambda^8 - \lambda^{6.4} \quad (kPa) \qquad (6)$$

The boundary condition at the outer wall surface corresponds to variation of transpulmonary pressure with time. Transpulmonary pressure is calculated by difference between alveolar pressure and pleural pressure Ptp=Palv-Ppl founded on physiology handbooks [13] as shown in Figure 2.

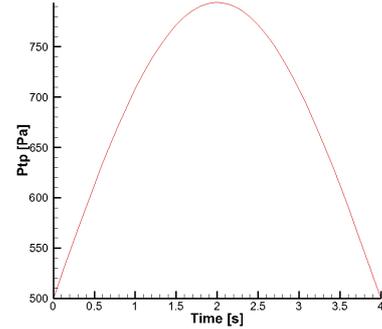

Figure 2: Transpulmonary pressure during normal breathing

The fluid and solid domains are coupled by satisfying the following three boundary conditions at the interface between the two domains: (i) displacements of the fluid and solid domain FSI boundaries must be compatible, (ii) tractions at FSI boundaries must be balanced and (iii) fluid obeys the no-slip condition. These conditions are given by Eqs 7-9.

$$d_i^f = d_i^s, \qquad (7)$$

$$n_j \sigma_{ij}^f = n_j \sigma_{ij}^s \quad where \quad \sigma_{ij}^f = -p\delta_{ij} + \mu(\frac{\partial v_i}{\partial x_j} + \frac{\partial v_j}{\partial x_i}), \qquad (8)$$

$$n_i v_i = n_i \frac{\partial d_i^s}{\partial t}, \qquad (9)$$

where $d_i^f$ and $d_i^s$ are the fluid and solid nodal displacement, $n_j$ is the interface normal vector, $\delta_{ij}$ is the Kronecker delta function, and $\sigma_{ij}^f$ and $\sigma_{ij}^s$ are the fluid and solid stress tensors.

**Fluid-structure interaction (FSI) implementation**
The software ADINA 8.7 (ADINA R&D) was utilized for our time-dependent simulation of fluid-structure interaction (FSI) between alveolar wall and lumen.
ADINA FSI offers two different methods, iterative and direct FSI Coupling, to solve the coupling between the fluid and the structural models. We have used directive method which is more applicable to solve large deformations with soft materials.

The Finite Element Method (FEM) is used to solve the governing equations, which discretizes the computational domain into finite elements that are interconnected by nodal points. In this work we make use of linear hexahedral, eight-node elements to discretize the fluid and solid domains in ADINA. The model is composed of 15087 hexahedral elements for the fluid and 22877 hexahedral elements for the solid domains. Mesh sensitivity analyses were conducted with four additional mesh sizes ranging from 14147 to 27,856 fluid elements and 14,000 to 46,495 solid elements. Independence in mesh size was obtained for the primary variables (velocity components, fluid pressure and structural displacements) within 5% relative error for the 4$^{th}$ and 3$^{th}$ mesh.

**Results and Discussion**
We have simulated normal breathing for different type of materials. In each case we compute inlet flux of alveolus during time. Figure 3 shows the results.
Integration of this curve gives us the expansion volume which can be used to obtain excursion ratio.

Edward et al. [14] reported 15.6% expansion ratio for the normal breathing. According to Tabel 1, our viscoelastic model gives closer results to their simulations.

Table 1: Results of excursion ratio for different material

| Material | Excursion Ratio |
| --- | --- |
| Elastic 4 kPa | 48.2% |
| Elastic 6 kPa | 34.0% |
| Elastic 8 kPa | 27.6% |
| Non-Linear Elastic | 54.7% |
| Viscoelastic | 15.8% |

The difference between elastic and viscoelastic model can be seen in Figure 3. Elastic model produce a sinusoidal flux profile at the inlet but non-linear elastic and viscoelastic material give a leaning sinusoidal profile.

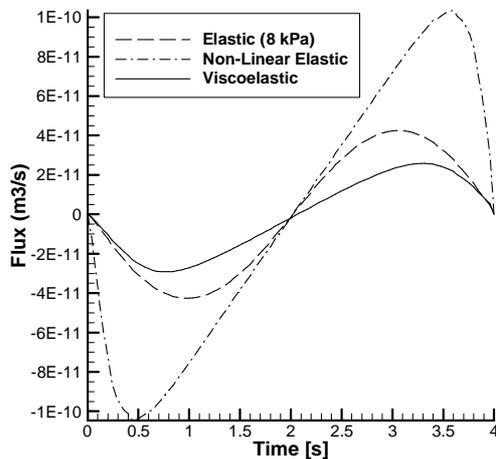
Figure 3: flux profile for different material models

As shown in Figure 3, the air flux at time 0.8 s is maximized for viscoelastic wall. The velocity contours and vectors for this time are shown in Figures 4 and 5. As shown, Velocity vectors show a laminar flow with no circulation, because of very low velocities at the end alveolus.

Distribution of effective stress in alveolar tissue can be seen in Figure 6. Maximum stress in this situation is 37 Pa.

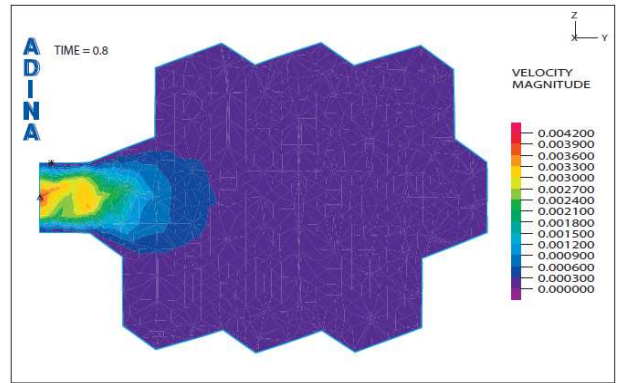
Figure 4: air velocity contours at t=0.8

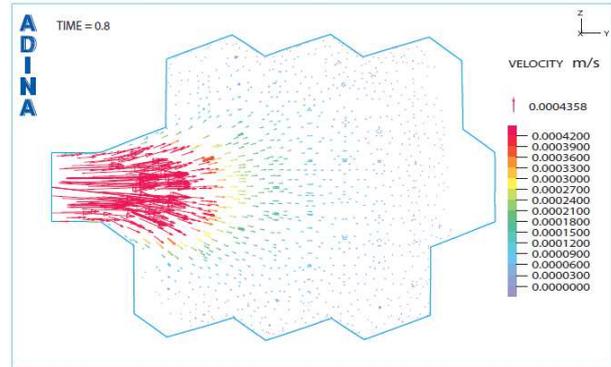
Figure 5: velocity vectors at t=0. 8

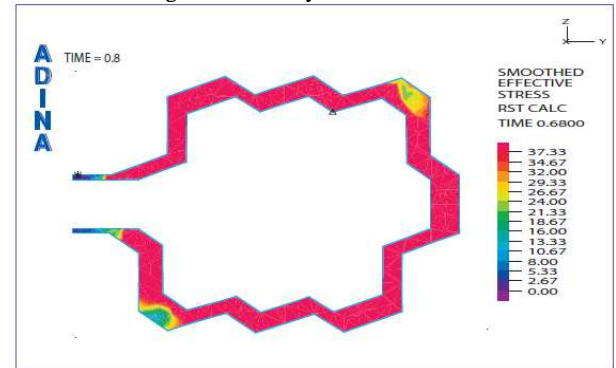
Figure 6: stress contours at t=0. 8

**Conclusions**
A fluid-structure analysis is performed to study normal breathing airflow. The effect of viscoelastic behavior of alveolar tissue on the profile of flow rate has been investigated. Linear elastic model presents a symmetric sinusoidal profile during breathing but in non-linear elastic and viscoelastic model maximum flow rate doesn't occur at the time of 1s. In addition expansion ratio in these models has considerable difference which among them viscoelastic model introduces the value of 12.4% which has the best agreement with other literatures.

**List of Symbols**

| | |
| --- | --- |
| $d_{f,s}\ d_{f,s}$ | Displacement vector (f denotes fluid, s denotes solid) |
| $E$ | Modulus of elasticity |
| $G$ | Reduced relaxation function |
| $n_i$ | Normal vector |
| $p$ | Fluid pressure |
| $P_{alv}$ | Alveolar pressure |
| $P_{pl}$ | Pleural pressure |
| $P_{tp}$ | Transpulmonary pressure |
| $v_i$ | Velocity vector |

**Greek symbols**

| | |
|---|---|
| $\delta_{ij}$ | Kronecker delta |
| $\varepsilon$ | Strain rate |
| $\lambda$ | Stretch ratio |
| $\mu$ | Dynamic viscosity of the fluid |
| $\sigma_0$ | Instantaneous stress response |
| $\sigma_{ij}^{f,s}$ | Cauchy stresses tensor (f denotes fluid, s denotes solid) |
| $\rho_{f,s}$ | Material density (f denotes fluid, s denotes solid) |